\documentclass[14pt]{extarticle}
\usepackage{amsmath}
\usepackage{amssymb}
\usepackage{bm}
\usepackage{color}
\usepackage{braket}
\usepackage{graphicx}
\usepackage{float}
\usepackage{caption}

\begin{document}
\begin{center}
\begin{Large}
{\bf Geometric measure of entanglement of quantum graph states prepared with controlled phase shift operators}
\end{Large}
\end{center}

\centerline {N. A. Susulovska \footnote{E-Mail address: n.a.susulovska@gmail.com}}
\medskip
\centerline {\small \it Ivan Franko National University of Lviv,}
\centerline {\small \it Professor Ivan Vakarchuk Department for Theoretical Physics,}
\centerline {\small \it 12 Drahomanov St., Lviv, 79005, Ukraine}

\abstract{ We consider graph states generated by the action of controlled phase shift operators on a separable state of a multi-qubit system. The case when all the qubits are initially prepared in arbitrary states is investigated. We obtain the geometric measure of entanglement of a qubit with the remaining system in graph states represented by arbitrary weighted graphs and establish its relationship with state parameters. For two-qubit graph states, the geometric measure of entanglement is also quantified on IBM's simulator \emph{Qiskit Aer} and quantum processor \emph{ibmq lima} based on auxiliary mean spin measurements. The results of quantum computations verify our analytical predictions.
}

\section{Introduction}

Over the past decades, a lot of effort has been directed towards theorizing and implementing a variety of practical schemes and algorithms, which could leverage the amazing potential of quantum mechanics (as an example see \cite{Bennet}, \cite{Bouwmeester}, \cite{Ekert}, \cite{Feynman}, \cite{Shor}, \cite{Montanaro}, \cite{Cerezo} and references therein). Studies of quantum entanglement, indisputably considered one of the fundamental quantum-mechanical features \cite{Einstein}, \cite{Horodecki}, took a central role in this endeavour and soon gave rise to quantum computing and quantum communications. Manifesting itself in peculiar long-range correlations, which result in non-factorable states of composite systems, quantum entanglement is viewed as an indispensable resource in a range of applications. For instance, this phenomenon lies in the foundation of quantum teleportation \cite{Bennet}, \cite{Bouwmeester}, quantum cryptography \cite{Ekert} and allows the unprecedented capabilities of quantum computers \cite{Feynman}. These devices operate with superpositions of quantum states in high-dimensional Hilbert spaces and harness the power of entanglement to solve complex and often classically intractable computational problems \cite{Shor}, \cite{Montanaro}, \cite{Cerezo}. In this light, exploring entangled quantum states and their physical properties as well as the ways to efficiently prepare such states on a quantum computer is of crucial importance.
	
Recently, graph states have received a considerable amount of attention due to their high degree and persistence of entanglement \cite{Briegel}, \cite{Raussendorf}, \cite{Hein}, \cite{Guhne}. These multipartite quantum states are widely used in areas such as quantum error correction \cite{Schlingemann}, \cite{Bell}, \cite{Liao}, quantum metrology \cite{Shettell}, \cite{Tao}, and quantum machine learning \cite{Gao}. As for the latter, graph states appear, for instance, as a resource in the hybrid qGAN model, being produced by a single layer of a variational circuit constituting a quantum generator \cite{Zoufal}, \cite{Gnatenko_4}. Taking into account such broad applicability, quantifying the entanglement of graph states becomes an important task, which has been considered in a number of studies both analytically and on the basis of quantum computations (for instance, see \cite{Gnatenko_4}, \cite{Wang}, \cite{Mooney}, \cite{Gnatenko_2}, \cite{Vesperini_1}, \cite{Vesperini_2}, \cite{Gnatenko_3} and references therein). Different entanglement measures have been adopted for this purpose. For instance, in \cite{Wang} the authors utilized the definition of negativity and performed a computationally heavy state tomography procedure to detect the full entanglement of graph states associated with ring graphs. In another range of studies \cite{Gnatenko_4}, \cite{Gnatenko_2}, \cite{Vesperini_1}, \cite{Vesperini_2}, \cite{Gnatenko_3}, \cite{Kuzmak_1}, \cite{Kuzmak_2}, the entanglement of different multi-qubit states was determined as the distance between an entangled target state and the nearest separable state. This quantity, first introduced in \cite{Shimony}, is known as the geometric measure of entanglement and can be formally represented by the following expression

\begin{equation}\label{eq:1:1}
    E(\vert\psi\rangle)=\min_{\{\vert\psi_s\rangle\}}(1-|\langle\psi|\psi_s\rangle|^2),
\end{equation}
where $|\psi\rangle$, $\{|\psi_s\rangle\}$ denote an entangled target state and a set of separable states, respectively, $d_{FS}^2 (\vert\psi\rangle, \vert\psi_s\rangle)=1-|\langle\psi|\psi_s\rangle|^2$ is a squared Fubini-Study distance between $|\psi\rangle$ and $|\psi_s\rangle$. In \cite{Cocchiarella} similar geometric considerations were extended to derive a distance-based entanglement measure for hybrid systems of qudits.
	
In general, when estimating the entanglement on a quantum computer, selecting an entanglement measure, which can be directly connected to some easily measurable observable is a huge benefit. It was shown in \cite{Frydryszak} that in order to calculate the geometric measure of entanglement of a spin $1/2$, which can represent a qubit, with an arbitrary quantum system in a pure state $|\psi \rangle$ it is enough to obtain the mean value of this spin. This useful property is reflected in the following relation
\begin{equation}\label{eq:1:2}
    E(|\psi \rangle) = \frac{1}{2} (1 - |\langle\bm{\sigma}\rangle|),
\end{equation}
where 
	
\begin{equation}\label{eq:mean-spin}
    |\langle\bm{\sigma}\rangle| = |\langle \psi| \bm{\sigma} | \psi \rangle | = \left(\langle\sigma^x\rangle^2 + \langle\sigma^y\rangle^2 + \langle\sigma^z\rangle^2\right)^{1/2}.
\end{equation}
Here $\sigma^x$, $\sigma^y$, $\sigma^z$ are Pauli operators.

Similarly, the author of \cite{Deb} determined that von Neumann entanglement entropy of the partial traces in bipartite systems of two-level atoms can be obtained based solely on the mean spin value corresponding to one of the atoms.
	
It should be stressed that the mean value of spin in an arbitrary pure state can be straightforwardly measured on a quantum computer according to the protocol described in \cite{Kuzmak_1}. Therefore, the quantity of entanglement associated with the corresponding quantum state can be estimated on the basis of such measurements. In recent years, many studies have exploited this idea to detect the geometric measure of entanglement of graph states with different structures. For instance, in \cite{Gnatenko_2} evolutionary graph states of spin systems with Ising interaction were considered. The authors of \cite{Vesperini_2} studied graph states generated by the action of controlled phase shift operators on the initial multi-qubit state corresponding to the uniform superposition over the computational basis. Paper \cite{Gnatenko_3} also focused on graph states prepared with the help of controlled phase shift operators, in this case, starting from the state of the system, in which all of the qubits are in arbitrary identical states. These studies concluded that the geometric measure of entanglement of an arbitrary qubit with other qubits in respective graph states depends on the degree of the corresponding graph vertex. This important result established a connection between a physical property of the quantum state and a geometric property of the graph used to describe it.
	
In the present study, we elaborate on the findings presented in \cite{Gnatenko_3} and revisit the problem of entanglement quantification for a class of multi-qubit graph states generated by the action of controlled phase shift operators. We examine a more general case, when the system of qubits is initially prepared in an arbitrary separable state. In addition, parameters of controlled phase shift operators corresponding to different graph edges take independent arbitrary values. This means that weighted graphs are used to represent quantum states under investigation. As a result of analytical considerations, a general expression for the geometric measure of entanglement of an arbitrary qubit with the rest of the system in a state corresponding to an arbitrary weighted graph is derived. We examine how this quantity depends on the initial state of the multi-qubit system as well as the parameterization of the entangling controlled phase shift operators. Furthermore, the geometric measure of entanglement for a selection of graph states is quantified both on IBM's quantum simulator \emph{Qiskit Aer} and real quantum backend \emph{ibmq lima}.
	
The paper is organized as follows. Section 2 is devoted to the analytical derivation of the geometric measure of entanglement of a qubit with other qubits in a graph state associated with an arbitrary weighted graph and its subsequent analysis. In Section 3 we detect the geometric measure of entanglement of two-qubit graph states on IBM's simulator \emph{Qiskit Aer} and quantum computer \emph{ibmq lima} \cite{IBM} and discuss the obtained results. Conclusions are presented in Section 4.
	
\section{Analytical consideration of the geometric measure of entanglement of graph states}
Consider an arbitrary multi-qubit state characterized by a general structure
	
\begin{equation}\label{eq:2:1}
    |\psi_G \rangle = \prod_{(i,j)\in E} U_{ij} |\psi_{init} \rangle,
\end{equation}
where $|\psi_{init} \rangle$ is an initial separable state of the system and $U_{ij}$ represents a two-qubit entangling unitary acting on states of qubits $q_i$, $q_j$. We can establish a one-to-one mapping between the class of quantum states (\ref{eq:2:1}) and a set of graphs $G(V, E)$. In this context, the set of graph vertices $V$ represents qubits, whereas the set of graph edges $E$ is associated with two-qubit operators acting on their initial states. Each unitary operator $U_{ij}$ can be written in an exponential form
	
\begin{equation}\label{eq:2:2}
    U_{ij} = \textrm{e}^{\textrm{i}\phi_{ij}H_{ij}},
\end{equation}
where $H_{ij}$ is a Hermitian operator, $\phi_{ij}$ is a scalar parameter.
	
In the present study, let us begin by considering a system of N qubits in the initial separable state 
	
\begin{equation}\label{eq:2:3}
    |\psi_{init} \rangle = \prod_{k \in V}|\psi(\alpha_k,\theta_k) \rangle,
\end{equation}
where
	
\begin{equation}\label{eq:2:4}
    |\psi(\alpha_k,\theta_k)\rangle = \text{cos} \frac{\theta_k}{2} |0\rangle + \textrm{e}^{\textrm{i}\alpha_k} \text{sin} \frac{\theta_k}{2} |1\rangle
\end{equation}
is an arbitrary one-qubit state, $0 \leq \alpha_k < 2\pi$, $0 \leq \theta_k \leq \pi$, $k \in V=\{0,...,N-1\}$. Conveniently, state (\ref{eq:2:4}) can be prepared with the help of the parameterized rotation operators $RY(\theta_k)$, $RZ(\alpha_k)$ acting on state $|0\rangle$ (accurate to the phase factor)
	
\begin{equation}\label{eq:2:5}
    |\psi(\alpha_k,\theta_k)\rangle = \textrm{e}^{-\textrm{i} \frac{\alpha_k}{2}} RZ(\alpha_k) RY(\theta_k) |0\rangle,
\end{equation}
here $RY(\theta_k) = \textrm{exp}(-\textrm{i} \theta_k \sigma_k^y / 2)$ and $RZ(\alpha_k) = \textrm{exp}(-\textrm{i} \alpha_k \sigma_k^z / 2)$.
	
Subsequently, a graph state associated with a weighted graph of a predefined structure can be obtained by applying the controlled phase shift operator $CP_{ij}(\phi_{ij})$ to each pair of qubits $q_i$, $q_j$ represented by vertices linked with an edge of weight $\phi_{ij}$. This two-qubit operator is defined as $CP_{ij}(\phi_{ij}) = |0\rangle_{i} {}_{i}\langle 0| \otimes I_{j} + |1\rangle_{i} {}_{i}\langle 1| \otimes P_{j}(\phi_{ij}),$ where $I_{j}$ is an identity operator and $P_{j}(\phi_{ij}) = |0\rangle_{j} {}_{j}\langle 0| + \textrm{e}^{\textrm{i} \phi_{ij}} |1\rangle_{j} {}_{j}\langle 1|$ is a phase shift operator acting on qubit $q_j$, $0 \leq \phi_{ij} < 2\pi$, $(i, j) \in E$. 
	
Note that we deal with weighted graphs and consider the case when all the phase shift parameters $\phi_{ij}$, $(i,j) \in E$ take different values. The resulting graph state reads
	
\begin{equation}\label{eq:2:9}
    |\psi_G\rangle = \prod_{(i,j) \in E} CP_{ij}(\phi_{ij}) \prod_{k \in V} |\psi(\alpha_k,\theta_k)\rangle, 
\end{equation}
where $|\psi(\alpha_k,\theta_k)\rangle$ is given by (\ref{eq:2:4}) and $CP_{ij}(\phi_{ij})$ acts on qubits $q_i$, $q_j$ as a control and a target, respectively. 
	
Here we can resort to an exponential form of the controlled phase shift operator
\begin{equation}\label{eq:2:10}
CP_{ij}(\phi_{ij}) = \textrm{e}^{\frac{\textrm{i} \phi_{ij}}{4} (I_i - \sigma_i^z) (I_j - \sigma_j^z)}.
\end{equation}
One can notice that our choice of entangling two-qubit operators allows us to simulate Ising interaction in systems of many spins 1/2.
	
Let us analytically estimate the geometric measure of entanglement of an arbitrary qubit $q_l$ with the remaining system in state (\ref{eq:2:9}) described by an arbitrary weighted graph. Note that we essentially study bipartite entanglement with one of the subsystems being represented by qubit $q_l$ and the rest of the qubits constituting the second subsystem. As follows from (\ref{eq:1:2}), our objective can be achieved by calculating the mean value of the corresponding Pauli operator $\langle \bm{\sigma}_l \rangle$ in the graph state. Namely, one has to separately consider $\langle \sigma^x_l \rangle$, $\langle \sigma^y_l \rangle$, $\langle \sigma^z_l \rangle$. Hereafter, we use notation $\langle ... \rangle = \langle \psi_G | ... | \psi_G \rangle$. 
	
Taking into account (\ref{eq:2:5}) yields
	
\begin{multline}\label{eq:2:11}
\langle \sigma^x_l \rangle = \langle\psi_0| \prod_{q\in V}  \textrm{e}^{\textrm{i}\frac{\theta_q}{2} \sigma^y_q} \textrm{e}^{\textrm{i}\frac{ \alpha_q}{2} \sigma^z_q} \prod_{(j,k) \in E} CP^{+}_{jk}(\phi_{jk}) \sigma^x_l \times \\ \times \prod_{(m,n) \in E} CP_{mn}(\phi_{mn}) \prod_{p \in V} \textrm{e}^{-\textrm{i}\frac{\alpha_p}{2} \sigma^z_p} \textrm{e}^{-\textrm{i}\frac{\theta_p}{2} \sigma^y_p} |\psi_0\rangle,
\end{multline}
where $|\psi_0\rangle = |0\rangle^{\otimes V}$. To simplify this expression we make use of identity (\ref{eq:2:10}) and take into consideration that Pauli operators $\sigma^x_l$ and $\sigma^z_l$ anticommute, thus obtaining
	
\begin{multline}\label{eq:2:12}
\prod_{(j,k) \in E} CP^{+}_{jk}(\phi_{jk}) \sigma^x_l \prod_{(m,n) \in E} CP_{mn}(\phi_{mn}) = \\ =\prod_{(j,k) \in E} \textrm{e}^{-\textrm{i}\frac{ \phi_{jk}}{4} (I_j- \sigma^z_j) (I_k - \sigma^z_k)} \: \sigma^x_l \prod_{(m,n) \in E} \textrm{e}^{\textrm{i}\frac{ \phi_{mn}}{4} (I_m-\sigma^z_m) (I_n - \sigma^z_n)}= \\
= \textrm{e}^{\frac{\textrm{i}}{2} \sigma^z_l \sum_{j \in N_G(l)} \phi_{jl} (I_j - \sigma^z_j)}  \sigma^x_l,
\end{multline}
where $N_G(l)$ denotes a set of vertices adjacent to the vertex $l$, known as its neighbourhood. Eventually, we find
	
\begin{equation}\label{eq:2:13}
\langle \sigma^x_l \rangle = \sin \theta_l \, \textrm{Re}\,z.
\end{equation}
Here $z \in \mathbb{C}$ reads
	
\begin{equation}\label{eq:2:14}
z = \text{e}^{-\text{i}\left(\alpha_l + \frac{1}{2} \sum_{j \in N_G(l)} \phi_{jl}\right)} \prod_{k \in N_G(l)} \left(\cos \frac{\phi_{kl}}{2} + \text{i} \; \sin \frac{\phi_{kl}}{2} \cos \theta_k \right),
\end{equation}
where $\sum_{j \in N_G(l)} \phi_{jl}$ is a weighted degree of the vertex denoting qubit $q_l$ in the graph.
	
Performing similar mathematical transformations we easily obtain the result for mean value $\langle \sigma^y_l \rangle$
	
\begin{equation}\label{eq:2:15}
\langle \sigma^y_l \rangle = -\sin \theta_l \, \textrm{Im}\,z,
\end{equation}
where $z$ is given by the same expression (\ref{eq:2:14}).
	
Lastly,
	
\begin{equation}\label{eq:2:16}
\langle \sigma^z_l \rangle = \langle \psi_0| \textrm{e}^{\textrm{i} \frac{\theta_l}{2} \sigma^y_l} \; \sigma^z_l  \textrm{e}^{-\textrm{i} \frac{\theta_l}{2} \sigma^y_l} |\psi_0\rangle = \cos \theta_l.
\end{equation}
	
Eventually, to find the geometric measure of entanglement of qubit $q_l$ with the rest of the qubits in state (\ref{eq:2:9}), we substitute mean values (\ref{eq:2:13}), (\ref{eq:2:15}), (\ref{eq:2:16}) into central expression (\ref{eq:1:2}) and obtain

\begin{equation}\label{eq:2:17}
    E_l (| \psi_G \rangle) = \frac{1}{2} \Biggl(1 - \biggl[ \sin^2 \theta_l \prod_{k \in N_G(l)} \Bigl(\cos^2 \frac{\phi_{kl}}{2} + \sin^2 \frac{\phi_{kl}}{2} \cos^2 \theta_k \Bigr) + \cos^2 \theta_l \vphantom{\prod_{k \in N_G(l)}}\biggr]^{1/2} \Biggr).
\end{equation}
As evident from (\ref{eq:2:17}), this entanglement measure depends on absolute values of a subset of parameters $\{\theta_m\}$  defining the initial state of the multi-qubit system, which corresponds to a closed neighbourhood $N_G[l]$ of vertex $l$ representing the qubit under consideration  (vertex $l$ itself combined with a set of its adjacent vertices), $m \in N_G[l]$. It also depends on absolute values of a set of parameters $\{\phi_{kl}\}$ passed to the controlled phase shift operators responsible for the generation of edges incident to vertex $l$, $k \in N_G(l)$.

Consider a special case of graph states (\ref{eq:2:9}) when all the qubits of the system are initially prepared in identical states (\ref{eq:2:4}) so that $\theta_m=\theta$, $m \in N_G[l]$. In addition assume that all the controlled phase shift operators associated with graph edges share the same parameter, hence $\phi_{kl}=\phi$, $k \in N_G(l)$. Under these constraints (\ref{eq:2:17}) is reduced to

\begin{equation}\label{eq:2:ent:special}
    E_l (| \psi_G \rangle) = \frac{1}{2} \Biggl(1 - \biggl[ \sin^2 \theta \Bigl(\cos^2 \frac{\phi}{2} + \sin^2 \frac{\phi}{2} \cos^2 \theta \Bigr)^{n_l} + \cos^2 \theta \biggr]^{1/2} \Biggr),
\end{equation}
which coincides with an expression obtained in \cite{Gnatenko_3}. Therefore, the results of the present study generalize our previous findings. Note, that in (\ref{eq:2:ent:special}) the geometric measure of entanglement of qubit $q_l$ explicitly depends on the degree of the corresponding graph vertex $n_l$, which is equal to the number of vertices in its neighbourhood. 

\section{Investigating the relation of the geometric measure of entanglement of graph states to the mean spin on a quantum device}

\begin{figure}[h!]
    \centering
    \vskip1mm
    \includegraphics[scale=0.6]{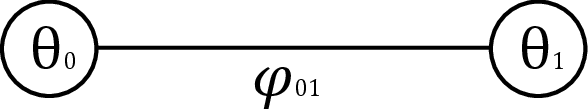}
    \caption{A graph representing state (\ref{eq:3:1}).}\label{fig:1}
\end{figure}

In order to put our theoretical findings to the test we examine two-qubit graph states of structure (\ref{eq:2:9}) on a quantum device and quantify their geometric measure of entanglement. Such states can be explicitly written as 

\begin{multline}\label{eq:3:1}
    |\psi_{G_2}\rangle = CP_{01}(\phi_{01}) | \psi (0, \theta_0) \rangle | \psi (0, \theta_1) \rangle = \\ = CP_{01}(\phi_{01}) \left(\text{cos} \frac{\theta_0}{2} |0\rangle + \text{sin} \frac{\theta_0}{2} |1\rangle\right) \left( \text{cos} \frac{\theta_1}{2} |0\rangle+ \text{sin} \frac{\theta_1}{2} |1\rangle \right)
\end{multline}
and associated with a graph depicted in Figure \ref{fig:1}. Graph state (\ref{eq:3:1}) is determined by a set of three parameters, namely $\theta_0$, $\theta_1$ defining the initial states of qubits in the bipartite system, and $\phi_{01}$ corresponding to the entangling gate parameter and, therefore, to the weight of the graph edge. Note that we set relative phases of the initial one-qubit states $\alpha_0$, $\alpha_1$ in (\ref{eq:2:4}) equal to zero since they don't have any impact on the geometric measure of entanglement, according to expression (\ref{eq:2:17}).

\begin{figure}[ht!] 
    \centering
    \vskip1mm
    \includegraphics[scale=1]{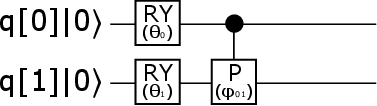}
    \caption{Quantum protocol for preparing two-qubit graph states (\ref{eq:3:1}).}\label{fig:2}
\end{figure}

On the basis of (\ref{eq:2:5}), we further obtain

\begin{equation}\label{eq:3:2}
    |\psi_{G_2}\rangle = CP_{01}(\phi_{01}) \textrm{e}^{-\frac{\textrm{i} \theta_1}{2} \sigma_1^y} \textrm{e}^{-\frac{\textrm{i} \theta_0}{2} \sigma_0^y} |00\rangle = CP_{01}(\phi_{01}) RY_1(\theta_1) RY_0(\theta_0) |00\rangle,
\end{equation}
which shows that an arbitrary two-qubit graph state can be prepared by the consecutive action of two rotation $RY$ gates and the controlled phase shift gate on the traditional initial state of a quantum register $|00\rangle$. See the protocol for generating quantum states of this structure corresponding to different weighted graphs on a gate-based quantum computer in Figure \ref{fig:2}.

For the purposes of this research, we aim to analyze how the geometric measure of entanglement of one qubit with another in graph state (\ref{eq:3:1}) is influenced by the choice of the initial separable state of the bipartite system as well as the parameter of the entangling gate. Hence, two special cases are investigated in detail. 

Firstly, we consider a subclass of graph states (\ref{eq:3:1}) with both parameters $\theta_0$, $\theta_1$ equal to $\pi/2$ and track the dependence of the geometric measure of entanglement on the parameter of the controlled phase shift gate $\phi_{01}$ as it runs in the range $[0,2\pi)$. Note that in this case the initial one-qubit states coincide with state $|+\rangle = \left(|0\rangle + |1\rangle\right)/\sqrt{2}$ (an eigenstate of Pauli-$X$ operator corresponding to eigenvalue $1$) and can be prepared with the help of Hadamard operators $H$. We have

\begin{equation}\label{eq:3:3}
    |\psi_{G_2} (\phi_{01})\rangle = CP_{01}(\phi_{01}) H_1 H_0 |00\rangle.
\end{equation}
According to (\ref{eq:2:17}), the geometric measure of entanglement of one qubit with another in state (\ref{eq:3:3}) is reduced to

\begin{equation}\label{eq:3:ent:1}
    E(|\psi_{G_2}(\phi_{01})\rangle) = \frac{1}{2} \left(1-\left|\cos \frac{\phi_{01}}{2}\right|\right).
\end{equation}

For the second subclass of graph states we fix parameter $\phi_{01}$ equal to $\pi$. Since $CZ = CP(\pi)$, in practice this means that the controlled-$Z$ operators $CZ$ are used to generate graph edges and the resulting states take the form

\begin{equation}\label{eq:3:4}
    |\psi_{G_2} (\theta_0, \theta_1)\rangle = CZ_{01} RY_1(\theta_1) RY_0(\theta_0) |00\rangle.
\end{equation}
In this setting we can showcase the dependency of the geometric measure of entanglement on the initial state parameters by letting them variate independently in the range $[0,\pi]$. Analytically, (\ref{eq:2:17}) yields

\begin{equation}\label{eq:3:ent:2}
    E(|\psi_{G_2}(\theta_0, \theta_1)\rangle) = \frac{1}{2} \left(1 - \left[\cos^2\theta_0 + \cos^2\theta_1 - \cos^2\theta_0 \cos^2\theta_1\right]^{1/2}\right).
\end{equation}

Expression (\ref{eq:1:2}) suggests that the geometric measure of entanglement in multi-qubit systems can be detected on a quantum device through the measurements of mean spin. Namely, one has to obtain mean values of Pauli operators $\sigma^x$, $\sigma^y$, $\sigma^z$ on the basis of quantum computations. This can be achieved, for instance, by following the approach presented in \cite{Kuzmak_1}. In our particular case, assume that we would like to estimate the geometric measure of entanglement of qubit $q_0$ in a certain graph state of a structure (\ref{eq:3:1}). It was shown that  $\langle \sigma^x_0 \rangle$, $\langle \sigma^y_0 \rangle$, $\langle \sigma^z_0 \rangle$ can be expressed in terms of probabilities defining the results of measurements performed on qubit $q_0$ in the computational basis $\{|0\rangle,|1\rangle\}$. We have

\begin{equation}
    \langle \sigma^x_0 \rangle = \langle \psi_{G_2} | \sigma^x_0 | \psi_{G_2} \rangle = \langle \tilde{\psi}^y_{G_2}| \sigma^z_0 | \tilde{\psi}^y_{G_2}\rangle = |\langle \tilde{\psi}^y_{G_2}| 0\rangle|^2 - |\langle \tilde{\psi}^y_{G_2}| 1\rangle|^2, \label{eq:3:5}
\end{equation}

\begin{equation}
    \langle \sigma^y_0 \rangle = \langle \psi_{G_2} | \sigma^y_0 | \psi_{G_2} \rangle = \langle \tilde{\psi}^x_{G_2}| \sigma^z_0 | \tilde{\psi}^x_{G_2} \rangle = |\langle \tilde{\psi}^x_{G_2}| 0\rangle|^2 - |\langle \tilde{\psi}^x_{G_2}| 1\rangle|^2, \label{eq:3:6}
\end{equation}

\begin{equation}
    \langle \sigma^z_0 \rangle = \langle \psi_{G_2} | \sigma^z_0 | \psi_{G_2} \rangle = |\langle \psi_{G_2}| 0\rangle|^2 - |\langle \psi_{G_2}| 1\rangle|^2, \label{eq:3:7}
\end{equation}
where 

\begin{align}
    |\tilde{\psi}^x_{G_2} \rangle &= \textrm{e}^{-\textrm{i} \pi \sigma^x_0/4} |\psi_{G_2}\rangle = RX_0(\pi/2) |\psi_{G_2}\rangle, \label{eq:3:8}\\
    |\tilde{\psi}^y_{G_2} \rangle &= \textrm{e}^{\textrm{i} \pi \sigma^y_0/4} |\psi_{G_2}\rangle = RY_0(-\pi/2)|\psi_{G_2}\rangle. \label{eq:3:9}
\end{align}
According to identities (\ref{eq:3:8}), (\ref{eq:3:9}), in order to detect mean values $\langle \sigma^x_0 \rangle$, $\langle \sigma^y_0 \rangle$ the state of qubit $q_0$ has to be rotated by angle $\pi/2$ around $y$ and $x$ axes, respectively, prior to the measurements in the computational basis.

\begin{figure}[ht!]
    \centering
    \vskip1mm
    \includegraphics[scale=0.5]{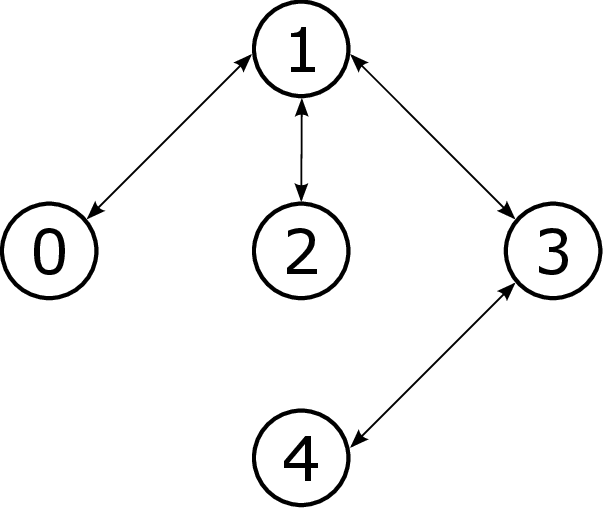}
    \caption{Architecture of IBM's quantum computer \emph{ibmq lima}. Arrows connect the qubits, to which CNOT gates can be directly applied. Each qubit can play a role of both a control and a target.}\label{fig:3}
\end{figure}

In the present study, we prepare graph states (\ref{eq:3:3}), (\ref{eq:3:4}) with the help of quantum circuits generalized in Figure \ref{fig:2} and estimate their entanglement according to the protocol described above on both IBM's simulator $Qiskit\;Aer$ and real quantum backend \emph{ibmq lima} \cite{IBM}. The latter is a universal gate-based quantum processor consisting of 5 superconducting qubits connected according to the map in Figure \ref{fig:3}.

Note that when preparing an arbitrary two-qubit state on a 5-qubit quantum computer, one has multiple choices of which physical qubits to utilize in the appropriate quantum circuit. Therefore, at the time of experiments calibration parameters of \emph{ibmq lima} including the readout error as well as one- and two-qubit gate errors were taken into account to minimize the cumulative error of computations.

\begin{figure}[ht!]
    \centering
    \includegraphics[scale=0.7]{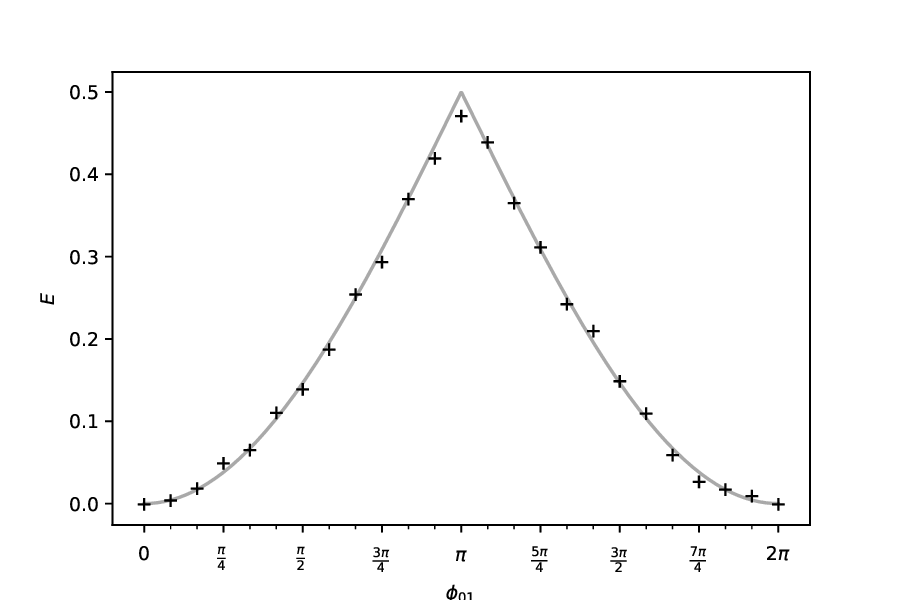}
    \caption{Results of quantifying the geometric measure of entanglement of qubit $q_0$ with qubit $q_1$ in graph state (\ref{eq:3:3}) on IBM's simulator \emph{Qiskit Aer} (marked with crosses) and analytical results (represented with a line).}\label{fig:4}
\end{figure}

\begin{figure}[ht!]
    \centering
    \includegraphics[scale=0.7]{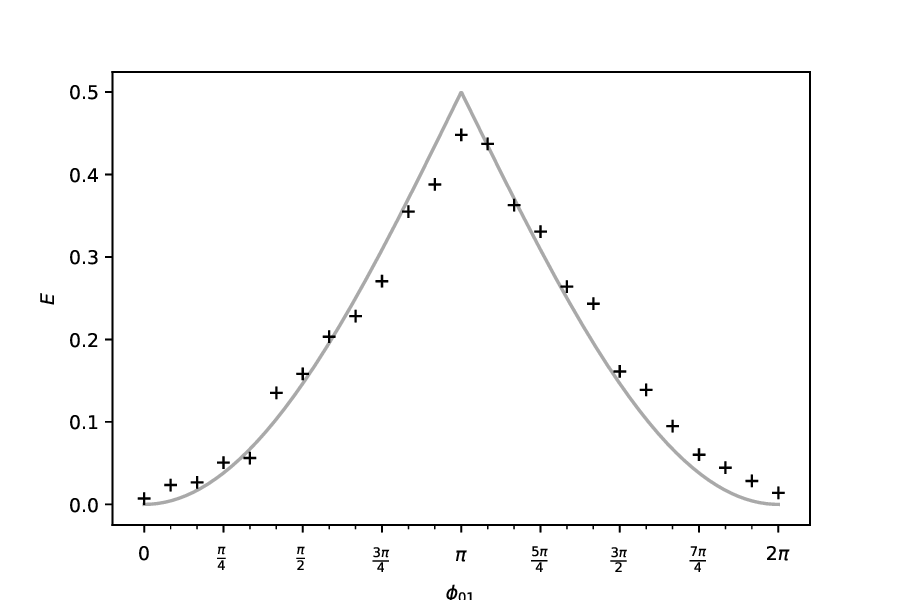}
    \caption{Results of quantifying the geometric measure of entanglement of qubit $q_0$ with qubit $q_1$ in graph state (\ref{eq:3:3}) on IBM's quantum computer \emph{ibmq lima} (marked with crosses) and analytical results (represented with a line).}\label{fig:5}
\end{figure}

To begin with, we quantify the geometric measure of entanglement of qubit $q_0$ with qubit $q_1$ in graph state $|\psi_{G_2}(\phi_{01})\rangle$ for different values of the controlled phase shift gate parameter $\phi_{01} \in [0, 2\pi]$ (see Figures \ref{fig:4}, \ref{fig:5}). The obtained experimental dependency is then compared to the analytical one on the basis of expression (\ref{eq:3:ent:1}). It's easy to see that on the given interval the geometric measure of entanglement reaches its maximum at $\phi_{01} = \pi$ and the respective experimental value is close to the theoretical prediction of 1/2. Note that this point on the plot corresponds to the graph state prepared by the action of the controlled-Z gate CZ

\begin{equation}\label{eq:3:maxent}
    |\psi_{G_2}\rangle = CZ_{01} H_1 H_0 |00\rangle.
\end{equation}
Hereafter, the slight quantitative misalignment between the results of quantum computations and the theoretical ones can be explained by the errors inherent to the quantum hardware that were mentioned before. The geometric measure of entanglement takes its minimal value of 0 at points $\phi_{01}=0, \, 2\pi$, which corresponds to the case when the controlled phase shift gate $CP$ is reduced to the identity gate $I$ and the resulting two-qubit state is separable.

\begin{figure}[ht!]
    \centering
    \includegraphics[scale=1]{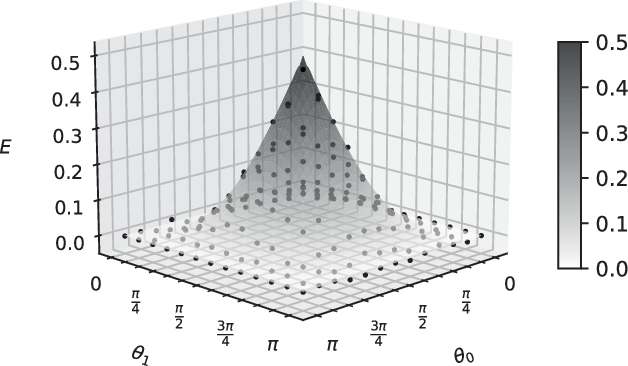}
    \caption{Results of quantifying the geometric measure of entanglement of qubit $q_0$ with qubit $q_1$ in graph state (\ref{eq:3:4}) on IBM's simulator \emph{Qiskit Aer} (marked with dots) and analytical results (represented with a continuous surface).}\label{fig:6}
\end{figure}

\begin{figure}[ht!]
    \centering
    \includegraphics[scale=1]{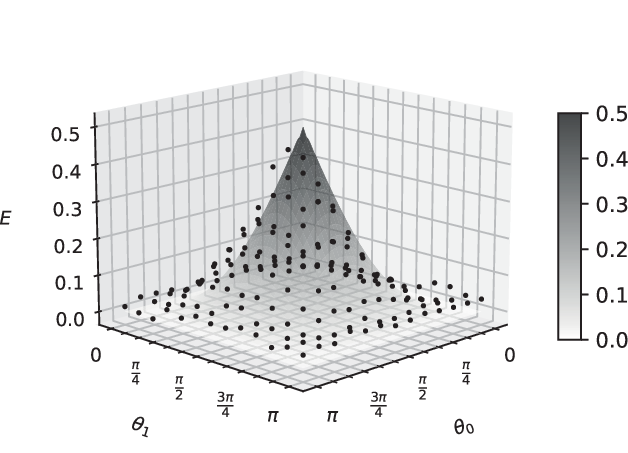}
    \caption{Results of quantifying the geometric measure of entanglement of qubit $q_0$ with qubit $q_1$ in graph state (\ref{eq:3:4}) on IBM's quantum computer \emph{ibmq lima} (marked with dots) and analytical results (represented with a continuous surface).}\label{fig:7}
\end{figure}

Similarly, the geometric measure of entanglement of qubit $q_0$ in graph state $|\psi_{G_2}(\theta_0, \theta_1) \rangle$ is obtained for different values of the initial state parameters $\theta_0, \theta_1 \in [0, \pi]$ (see Figures \ref{fig:6}, \ref{fig:7}). We show that the results of quantum computations are consistent with previously derived analytical expression (\ref{eq:3:ent:2}). In particular, graph state (\ref{eq:3:4}) is maximally entangled when both of the initial state parameters $\theta_0$, $\theta_1$ are equal to $\pi/2$. Considering that $RY(\pi/2)= X H$, and state $|+\rangle$ is an eigenstate of the operator Pauli-$X$ associated with eigenvalue $1$, this brings us back to the state (\ref{eq:3:maxent}) as expected. On the contrary, the graph state under consideration becomes separable (the geometric measure of entanglement goes to 0) if at least one of the initial state parameters $\theta_0$, $\theta_1$ is set equal to $0$ or $\pi$. From the standpoint of quantum programming, to prepare such one-qubit states the rotation $RY$ gate is replaced in the quantum circuit by the identity gate $I$ and the Pauli-$Y$ gate (accurate to a phase factor), respectively.

As expected, in both Figures \ref{fig:4}, \ref{fig:5} and Figures \ref{fig:6}, \ref{fig:7} the results produced by a real quantum processor deviate from the theoretical results slightly more than the simulated ones due to the presence of noise. However, there is still a good agreement with our analytical predictions. Note that the associated computational errors are not significant since we only consider two-qubit systems and all the quantum circuits executed for the purposes of this study are fairly shallow.

\section{Conclusions}
In this paper a class of graph states (\ref{eq:2:9}) obtained as a result of the action of controlled phase shift operators on the initial separable state of a multi-qubit system, in which all of the qubits are in arbitrary states, has been considered. We have derived an analytical expression (\ref{eq:2:17}) for the geometric measure of entanglement of an arbitrary qubit with other qubits in a graph state belonging to this class, which is described by an arbitrary weighted graph. This expression shows that the geometric measure of entanglement is related to the subset of absolute values of initial state parameters corresponding to the closed neighbourhood of the vertex representing the qubit under consideration as well as the set of absolute values of parameters passed to the controlled phase shift operators responsible for generating its incident edges.

In addition, the entanglement of two-qubit graph states (\ref{eq:3:1}) has been studied on the basis of quantum computations. Namely, we have proposed a protocol for preparing such quantum states (see Figure \ref{fig:2}) and detected the geometric measure of entanglement associated with their two special cases through auxiliary mean spin measurements on IBM's quantum simulator \emph{Qiskit Aer} and quantum device \emph{ibmq lima}. In the first case,  both of the initial state parameters $\theta_0$, $\theta_1$ have been set equal to $\pi/2$ (\ref{eq:3:3}), which has allowed us to estimate the dependence of the entanglement on the parameter of the phase shift gate $\phi_{01}$. Alternatively, in the second case, we have generated graph states with parameter $\phi_{01}$ equal to $\pi$ (\ref{eq:3:4}) and analyzed how the choice of the initial separable bipartite state impacts the entanglement in the system. The results obtained on a quantum device in the course of this research are in a good agreement with analytical ones.

\section*{Acknowledgements}
The author thanks Prof. Gnatenko Kh. P. for useful comments and invaluable support during the research studies.


\begin{thebibliography}{99}
    
    \bibitem {Bennet} C. H. ~Bennett, G. ~Brassard, C. ~Cr\'epeau, R. ~Jozsa, A. ~Peres, W. K. ~Wootters. Teleporting an unknown quantum state via dual classical and Einstein-Podolsky-Rosen channels. \textit{Phys. Rev. Lett.} \textbf{70}, 1895 (1993) [DOI: https://doi.org/10.1103/PhysRevLett.70.1895].
    
    \bibitem {Bouwmeester} D. ~Bouwmeester, J.-W. ~Pan, K. ~Mattle, M. ~Eibl, H. ~Weinfurter, A. ~Zeilinger. Experimental quantum teleportation. \textit{Nature} \textbf{390}, 575 (1997) [DOI: https://doi.org/10.1038/37539].
    
    \bibitem {Ekert} A. K.~Ekert. Quantum cryptography based on Bell's theorem. \textit{Phys. Rev. Lett.} \textbf{67}, 661 (1991) [DOI: https://doi.org/10.1103/PhysRevLett.67.661].
    
    \bibitem {Feynman} R. P.~Feynman. Simulating physics with computers. \textit{Int. J. Theor. Phys.} \textbf{21}, 467 (1982) [DOI: https://doi.org/10.1007/BF02650179].
    
    \bibitem {Shor} P. W.~Shor. Polynomial-Time algorithms for prime factorization and discrete logarithms on a quantum computer. \textit{SIAM J. Comp.} \textbf{26}, 1484 (1997) [DOI: https://doi.org/10.1137/S0097539795293172].
    
    \bibitem {Montanaro} A.~Montanaro. Quantum algorithms: an overview. \textit{npj Quant. Inf.} \textbf{2}, 15023 (2016) [DOI: https://doi.org/10.1038/npjqi.2015.23].
    
    \bibitem {Cerezo} M.~Cerezo, A. ~Arrasmith, R. ~Babbush, S. C. ~Benjamin, S. ~Endo, K. ~Fujii, J. R. ~McClean, K. ~Mitarai, X. ~Yuan, L. ~Cincio, P. J. ~Coles. Variational quantum algorithms. \textit{Nat. Rev. Phys.} \textbf{3}, 625 (2021) [DOI: https://doi.org/10.1038/s42254-021-00348-9].
    
    \bibitem {Einstein} A.~Einstein, B.~Podolsky, N.~Rosen. Can quantum-mechanical description of physical reality be considered complete? \textit{Phys. Rev.} \textbf{47}, 777 (1935) [DOI: https://doi.org/10.1103/PhysRev.47.777].
    
    \bibitem {Horodecki} R.~Horodecki, P.~Horodecki, M.~Horodecki. Quantum entanglement. \textit{Rev. Mod. Phys.} \textbf{81}, 865 (2009) [DOI: https://doi.org/10.1103/RevModPhys.81.865].
    
    \bibitem {Briegel} H. J. ~Briegel, R.~Raussendorf. Persistent entanglement in arrays of interacting particles. \textit{Phys. Rev. Lett.} \textbf{86}, 910 (2001) [DOI: https://doi.org/10.1103/PhysRevLett.86.910].
    
    \bibitem {Raussendorf} R.~Raussendorf, H. J. ~Briegel. A one-way quantum computer. \textit{Phys. Rev. Lett.} \textbf{86}, 5188 (2001) [DOI: https://doi.org/10.1103/PhysRevLett.86.5188].
    
    \bibitem {Hein} M.~Hein, J.~Eisert, H. J. ~Briegel. Multiparty entanglement in graph states. \textit{Phys. Rev. A} \textbf{69}, 062311 (2004) [DOI: https://doi.org/10.1103/PhysRevA.69.062311].
    
    \bibitem {Guhne} O. ~G\"uhne, G. ~T\'oth, P. ~Hyllus, H. J. ~Briegel. Bell inequalities for graph states. \textit{Phys. Rev. Lett.} \textbf{95}, 120405 (2005) [DOI: https://doi.org/10.1103/PhysRevLett.95.120405].
    
    \bibitem {Schlingemann} D. ~Schlingemann, R. F. ~Werner. Quantum error-correcting codes associated with graphs. \textit{Phys. Rev. A} \textbf{65}, 012308 (2001) [DOI: https://doi.org/10.1103/PhysRevA.65.012308].
    
    \bibitem {Bell} B. A. ~Bell, D. A. ~Herrera-Mart\'i, M. S. ~Tame, D. ~Markham, W. J. ~Wadsworth, J. G. ~Rarity. Experimental demonstration of a graph state quantum error-correction code. \textit{Nat. Commun.} \textbf{5}, 3658 (2014) [DOI: https://doi.org/10.1038/ncomms4658].
    
    \bibitem {Liao} P. ~Liao, B. C. ~Sanders, D. L. ~Feder. Topological graph states and quantum error-correction codes. \textit{Phys. Rev. A} \textbf{105}, 042418 (2022) [DOI: https://doi.org/10.1103/PhysRevA.105.042418].
    
    \bibitem {Shettell} N. ~Shettell, D. ~Markham. Graph states as a resource for quantum metrology. \textit{Phys. Rev. Lett.} \textbf{124}, 110502 (2020) [DOI: https://doi.org/10.1103/PhysRevLett.124.110502].
    
    \bibitem {Tao} H. ~Tao, X. ~Tan. Quantum multiparameter estimation with graph states. arXiv:2306.02518 [quant-ph].
    
    \bibitem {Gao} X. ~Gao, Z.-Y. ~Zhang, L.-M. ~Duan. A quantum machine learning algorithm based on generative models. \textit{Sci. Adv.} \textbf{4}, 12 (2018) [DOI: 10.1126/sciadv.aat9004].
    
    \bibitem {Zoufal} C. ~Zoufal, A. ~Lucchi, S. ~Woerner. Quantum Generative Adversarial Networks for learning and loading random distributions. \textit{npj Quant. Inf.} \textbf{5}, 103 (2019) [DOI: https://doi.org/10.1038/s41534-019-0223-2].
    
    \bibitem {Gnatenko_4} Kh. P. ~Gnatenko. Evaluation of variational quantum states entanglement on a quantum computer by the mean value of spin. arXiv:2301.03885 [quant-ph].
    
    \bibitem {Wang} Y. ~Wang, Y. ~Li, Zq. ~Yin, B. Zeng. 16-qubit IBM universal quantum computer can be fully entangled. \textit{npj Quant. Inf.} \textbf{4}, 46 (2018) [DOI: https://doi.org/10.1038/s41534-018-0095-x].
    
    \bibitem {Mooney} G. J. ~Mooney, Ch. D. ~Hill, L. C. L. ~Hollenberg. Entanglement in a 20-qubit superconducting quantum computer. \textit{Sci. Rep.} \textbf{9}, 13465 (2019) [https://doi.org/10.1038/s41598-019-49805-7].
    
    \bibitem {Gnatenko_2} Kh. P. ~Gnatenko, V. M. ~Tkachuk. Entanglement of graph states of spin system with Ising interaction and its quantifying on IBM's quantum computer. \textit{Phys. Lett. A} \textbf{396}, 127248 (2021) [DOI: https://doi.org/10.1016/j.physleta.2021.127248].
    
    \bibitem {Vesperini_1} A. ~Vesperini. Correlations and projective measurements in maximally entangled multipartite states. \textit{Ann. Phys.} \textbf{457}, 169406 (2023) [DOI: https://doi.org/10.1016/j.aop.2023.169406].
    
    \bibitem {Vesperini_2} A. ~Vesperini, R. ~Franzosi. Entanglement, quantum correlators and connectivity in graph states. arXiv:2308.07690 [quant-ph].
    
    \bibitem {Gnatenko_3} Kh. P. ~Gnatenko, N. A. ~Susulovska. Geometric measure of entanglement of multi-qubit graph states and its detection on a quantum computer. \textit{EPL} \textbf{136}, 40003 (2022) [DOI: 10.1209/0295-5075/ac419b].
    
    \bibitem {Kuzmak_1} A. R. ~Kuzmak, V. M. ~Tkachuk. Detecting entanglement by the mean value of spin on a quantum computer. \textit{Phys. Lett. A} \textbf{384}, 126579 (2020) [DOI: https://doi.org/10.1016/j.physleta.2020.126579].
    
    \bibitem {Kuzmak_2} A. R. ~Kuzmak, V. M. ~Tkachuk. Preparation and study of the entanglement of the Schr\"odinger cat state on the ibmq-melbourne quantum computer. \textit{Condens. Matter Phys.} \textbf{23}, 43001 (2020) [DOI: 10.5488/CMP.23.43001].
    
    \bibitem {Shimony} A.~Shimony. Degree of entanglement. \textit{Ann. N.Y. Acad. Sci.} \textbf{755}, 675 (1995) [DOI:  https://doi.org/10.1111/j.1749-6632.1995.tb39008.x].
    
    \bibitem {Cocchiarella} D. ~Cocchiarella, S. ~Scali, S. ~Ribisi, B. ~Nardi, G. ~Bel-Hadj-Aissa, R. ~Franzosi. Entanglement distance for arbitrary $M$-qudit hybrid systems. \textit{Phys. Rev. A} \textbf{101}, 042129 (2020) [DOI: https://doi.org/10.1103/PhysRevA.101.042129].
    
    \bibitem {Frydryszak} A. M. ~Frydryszak, M. I. ~Samar, V. M. ~Tkachuk. Quantifying geometric measure of entanglement by mean value of spin and spin correlations with application to physical systems. \textit{Eur. Phys. J. D} \textbf{71}, 233 (2017) [DOI: https://doi.org/10.1140/epjd/e2017-70752-3].
    
    \bibitem {Deb} R. N.~Deb. Von Neumann entropy in a dispersive cavity. \textit{J. Mod. Opt.} \textbf{68}, 19 (2021) [DOI: 10.1080/09500340.2021.1970306].
    
    \bibitem{IBM} IBM Q experience. https://quantum-computing.ibm.com/

\end{thebibliography}
\end{document}